\DocumentMetadata{
	lang		= eng-US, 	
	pdfstandard	= A-3u,		
	pdfversion	= 1.7, 		
}

\documentclass[colorlinks,upint,subscriptcorrection,varvw,hyphenate,balance,greek,russian,vietnamese,german,nofoot]{asmeconf} 

\hypersetup{%
	pdfauthor={John H. Lienhard},									  
	pdftitle={ASME Conference Paper LaTeX Template},                  
	pdfkeywords={ASME conference paper, LaTeX template, BibTeX style},
	pdfsubject = {Describes the asmeconf LaTeX template},			  
}

\allowdisplaybreaks 

\usepackage{comment}
\begin{document}



%

\title{Graph Attention-Based Virtual Metrology for Film Deposition Processes in Semiconductor Manufacturing} 
 
%
%
%

\SetAuthors{%
    Tao Han\affil{1}\affil{2}\JointFirstAuthor, 
	Suk Ki Lee\affil{1}\JointFirstAuthor,  
	Hyunwoong Ko\affil{1}\CorrespondingAuthor{Hyunwoong.Ko@asu.edu}
	}

\SetAffiliation{1}{School of Manufacturing Systems and Networks, Arizona State University, Mesa, AZ}
\SetAffiliation{2}{Intel Foundry, OTF Site, AZ}



\maketitle





\begin{abstract}

Artificial intelligence-driven semiconductor manufacturing increasingly operates at nanometer and angstrom scales, where precise process control depends on accurate and timely metrology. However, physical metrology is inherently limited by measurement latency, cost, and sampling constraints, which restrict its scalability in high-volume production environments. Virtual metrology (VM) has therefore emerged as an effective alternative by predicting wafer-level characteristics from equipment sensor data. Despite recent advances, many existing VM models remain primarily correlation-driven and lack the ability to capture structured dependencies among heterogeneous process variables, while also providing limited interpretability.
This study presents a graph attention-based VM framework for film deposition processes that integrates temporal feature learning with structured parameter–layer dependency modeling. The proposed approach represents each process step–parameter pair as a node and extracts temporal embeddings from high-frequency equipment traces using convolutional feature encoders. A parameter-to-layer graph attention mechanism is then employed to model directional dependencies, enabling each film layer to selectively aggregate relevant process information.
The framework is evaluated using industrial deposition data collected from production wafers, where the model predicts film thickness from multivariate sensor signals. Experimental results demonstrate improved predictive performance compared to baseline models. In addition, analysis of the learned attention weights reveals interpretable parameter–layer relationships that are consistent with physical process behavior, capturing both dominant process factors and temporal dependencies across deposition stages.
These results indicate that the proposed framework not only enhances prediction accuracy but also provides meaningful insight into process dynamics, supporting more effective monitoring and optimization in semiconductor manufacturing.
\end{abstract}


\section{Introduction}
The rapid expansion of Artificial Intelligence (AI), high-performance computing, and large-scale digital infrastructures has dramatically increased the global demand for advanced semiconductor devices \cite{siliconminds}.  
Semiconductor manufacturing, therefore, plays a foundational role in enabling modern digital technologies, supporting applications ranging from data centers and intelligent systems to consumer electronics and communication networks \cite{developmenttrendsofsemicon}. 
As device architectures continue to evolve toward increasingly complex nanoscale structures, fabrication processes must achieve extremely high levels of precision, stability, and repeatability across hundreds of sequential manufacturing steps \cite{nanoscalechallenge, lee2026generative}. 
Within this complex production environment, thin film deposition represents one of the most critical unit operations because the physical and chemical properties of deposited films directly influence device performance, reliability, and yield \cite{ThinFilmreview}. 
Modern integrated circuits may contain dozens to hundreds of deposited material layers, including dielectric, conductive, and semiconductor films, each requiring stringent control of thickness, composition, and uniformity during fabrication \cite{integratedcircuitfabrication}. 

To ensure that such strict specifications are satisfied during manufacturing, semiconductor fabrication facilities rely heavily on advanced metrology techniques to measure film thickness and other structural characteristics throughout the production flow \cite{metrologyfornextgensemicon}. 
However, physical metrology measurements introduce additional process latency, measurement variability, and equipment capacity constraints in high-volume manufacturing environments \cite{metrorequirement}. 
As semiconductor production scales further in complexity and throughput, these limitations increasingly motivate the development of predictive approaches that can estimate wafer characteristics without performing direct physical measurements \cite{exploringMLpredict}. 
Virtual metrology (VM) has therefore emerged as an important paradigm that leverages equipment sensor data and statistical modeling to infer post-process wafer properties \cite{approachVM}. 

In recent years, machine learning (ML) approaches, including ensemble methods, deep neural networks, and temporal sequence models, have been widely explored to improve VM prediction accuracy by learning complex relationships between multivariate process signals and wafer-level outcomes \cite{exploringMLpredict}.
Particularly, in film deposition processes, VM models have demonstrated the potential to estimate film thickness using high-frequency equipment traces such as chamber pressure, temperature, and gas flow signals recorded during wafer processing \cite{VMdevelopmentforSIN}. 
However, many existing VM approaches remain primarily correlation-driven and do not explicitly account for structured causal relationships among heterogeneous process variables \cite{Review_AI_for_OP_Met}. 
Semiconductor deposition processes involve complex interactions among multiple equipment parameters across different processing stages, where the historical evolution of certain signals may influence downstream process outcomes \cite{semiconprocessfundamentals}. 
Conventional regression models or standard neural architectures may capture statistical associations but often struggle to represent such structured multivariate causal relationships in a principled manner \cite{Review_AI_for_OP_Met}. 
In addition, the lack of interpretability in many deep learning-based VM models limits their practical utility for diagnosing process deviations or identifying influential parameters in industrial manufacturing environments \cite{exploringMLpredict}. 

To address these challenges, this study proposes a graph attention-based VM framework for film deposition processes that integrates temporal feature learning with structured causal relationship modeling between process parameters and deposited film layers. 
The proposed method represents each process step–parameter pair as a node in a graph structure and extracts temporal embeddings from equipment trace signals using convolution-based feature encoders. 
A graph attention mechanism is then employed to model parameter-to-layer relationships, allowing each film layer representation to selectively aggregate relevant process information from multiple parameter nodes. 
Furthermore, the framework employs a parameter-to-layer graph attention mechanism to capture directional causal relationships between process parameters and film layers, improving predictive capability while providing the potential for process interpretability. 
The effectiveness of the proposed approach is evaluated using industrial deposition process data collected from production wafers, where the model predicts film thickness based on high-frequency equipment sensor measurements and demonstrates improved performance compared with baseline models.

The remainder of this paper is organized as follows.  
Section~2 reviews prior research on semiconductor manufacturing metrology and ML-based VM methods.  
Section~3 formulates the problem and presents the proposed graph attention and causal learning framework.  
Section~4 describes the experimental setup and evaluates the model using real industrial deposition data.  
Finally, Section~5 concludes the paper and discusses potential directions for future research.

\section{Related Work}

\subsection{Semiconductor Manufacturing Metrology}

Accurate metrology plays a critical role in semiconductor manufacturing by enabling process monitoring, feedback control, and yield optimization across complex fabrication workflows \cite{decisionVM}.  
As device geometries continue to shrink and process complexity increases, the ability to precisely measure structural and material properties such as film thickness, critical dimensions, and surface morphology becomes essential for maintaining manufacturing stability \cite{metrologyfornextgensemicon}.  
Modern semiconductor fabrication, therefore, relies on a diverse set of metrology techniques designed to characterize nanoscale features with high precision and throughput \cite{metrologyfornextgensemicon}.

Among these techniques, critical dimension scanning electron microscopy (CD-SEM) has been widely adopted for inline dimensional measurement due to its high spatial resolution and relatively high measurement throughput \cite{CDSEM}.  
Optical metrology methods, including spectroscopic ellipsometry and optical scatterometry, are also extensively used for thin-film characterization because they enable non-destructive measurement of film thickness and optical properties \cite{SE}.  
These optical approaches typically rely on inverse modeling techniques that match measured optical responses with simulated models to estimate structural parameters \cite{SE}.  
In addition to inline metrology tools, high-resolution characterization techniques such as transmission electron microscopy (TEM) and atom probe tomography (APT) are commonly used as reference measurements to investigate material structures at atomic or near-atomic resolution \cite{Review_AI_for_OP_Met}.

Despite their accuracy, physical metrology techniques face several challenges in high-volume semiconductor manufacturing environments \cite{Review_AI_for_OP_Met}.  
Measurement throughput limitations, equipment cost, and sampling constraints often prevent metrology tools from measuring every wafer or every process step \cite{metrology}.  
Furthermore, as feature sizes approach the atomic scale, measurement uncertainty and process-induced variations can become comparable in magnitude, making it increasingly difficult to maintain accurate and stable measurements across production wafers \cite{metrology}.  
These limitations have motivated the exploration of data-driven alternatives that can complement or partially replace conventional metrology approaches \cite{datadrivenmetro}.

\subsection{Machine Learning-Based Virtual Metrology}

VM has emerged as an important approach for predicting wafer-level quality metrics using equipment sensor data collected during semiconductor processing \cite{VMforsemiconmfg}.  
Instead of relying solely on direct physical measurements, VM models infer post-process characteristics by learning relationships between process variables and final wafer outcomes \cite{VMforsemiconmfg}.  
This capability enables predictive monitoring of manufacturing processes while reducing reliance on costly and time-consuming metrology operations \cite{Review_AI_for_OP_Met}.
Early VM studies primarily relied on statistical learning techniques such as linear regression, partial least squares regression, and support vector machines to model relationships between equipment parameters and wafer measurements \cite{VMforsemiconmfg}.  
While these models provided initial demonstrations of VM feasibility, their ability to capture complex nonlinear relationships in modern semiconductor processes remained limited \cite{metrologyfornextgensemicon}.

More recently, ML and deep learning (DL) techniques have been increasingly adopted to improve VM performance \cite{Review_AI_for_OP_Met}.  
Ensemble learning methods such as random forests and gradient boosting models have been applied to handle high-dimensional sensor data and nonlinear process relationships \cite{reviewregandpred}.  
More recently, DL approaches have significantly advanced VM performance. 
Architectures such as convolutional neural networks (CNNs) and recurrent neural networks (RNNs) have been widely used to capture temporal patterns in equipment sensor traces. 
More recently, advanced DL architectures have been developed that offer the potential to further improve VM performance. 
In particular, attention-based learning mechanisms have shown promise for modeling complex dependencies in high-dimensional process data \cite{cnnattentionforpred, lee2024amtransformer}. 
Graph neural networks provide a natural framework for representing structured relationships among process variables and capturing interactions between equipment parameters observed during wafer processing \cite{GNNVMPVD}. 
By combining graph-based representations with attention mechanisms, such models can selectively identify influential process signals and learn structured dependencies among parameters that contribute to wafer-level outcomes \cite{SukkiLee31122026}.
These models have demonstrated promising predictive accuracy for estimating wafer-level properties such as film thickness, etch depth, and critical dimensions \cite{SukkiLee31122026}.

In film deposition processes specifically, VM approaches have leveraged equipment trace signals such as chamber temperature, gas flow rates, and plasma emission spectra to estimate film thickness and related quality metrics \cite{VMforsemiconmfg}.  
Several studies have reported high prediction accuracy using sensor data collected during chemical vapor deposition or plasma processing steps, highlighting the potential of ML-driven VM for improving process monitoring and control \cite{VMdevelopmentforSIN}.

\subsection{Limitations of Existing Virtual Metrology Models}
Despite recent progress in ML-based VM, several challenges remain in applying these methods to complex semiconductor manufacturing processes \cite{VMforsemiconmfg}.  
One key limitation is that many existing VM models primarily rely on correlation-based learning and do not explicitly represent structured relationships among process variables \cite{GNNVMPVD}. 
Semiconductor fabrication processes often involve complex interactions among multiple equipment parameters across different processing stages, where the evolution of one variable may influence subsequent process outcomes \cite{multistageprocessdiag}.  
Existing VM ML models may capture statistical correlations between variables but often fail to represent these structured dependencies in a principled manner \cite{GNNVMPVD}.

Another important challenge lies in the interpretability of VM models \cite{lehnert2024xplainable}.
DL approaches frequently operate as black-box predictors, making it difficult for process engineers to understand which equipment parameters or process conditions are responsible for predicted variations in wafer characteristics \cite{reviewregandpred}.  
Such limitations reduce the usefulness of VM models for root cause analysis and process optimization in industrial manufacturing environments \cite{reviewregandpred}.
These challenges motivate the development of modeling approaches that can explicitly capture structured relationships among process parameters while also providing improved interpretability of model predictions \cite{SukkiLee31122026}. 
In this work, we address these limitations by introducing a graph attention-based VM framework that models parameter-to-layer dependencies and incorporates causal learning to better represent interactions among process variables at different process steps.


\section{Proposed Methodology}

This section presents the proposed graph attention-based VM framework for predicting film thickness in semiconductor deposition processes using equipment sensor data. 
The proposed model learns the relationship between multivariate process signals collected during wafer processing and the resulting film thickness measurements.
During wafer processing, multiple equipment parameters are recorded across different process steps, producing multivariate time-series signals that describe the evolution of chamber conditions. 
These signals capture dynamic process behaviors such as variations in precursor flow, chamber pressure, and temperature, which directly influence deposition kinetics and film growth mechanisms.

\begin{figure*}[h]
\centering
   \includegraphics[width=0.99\textwidth]{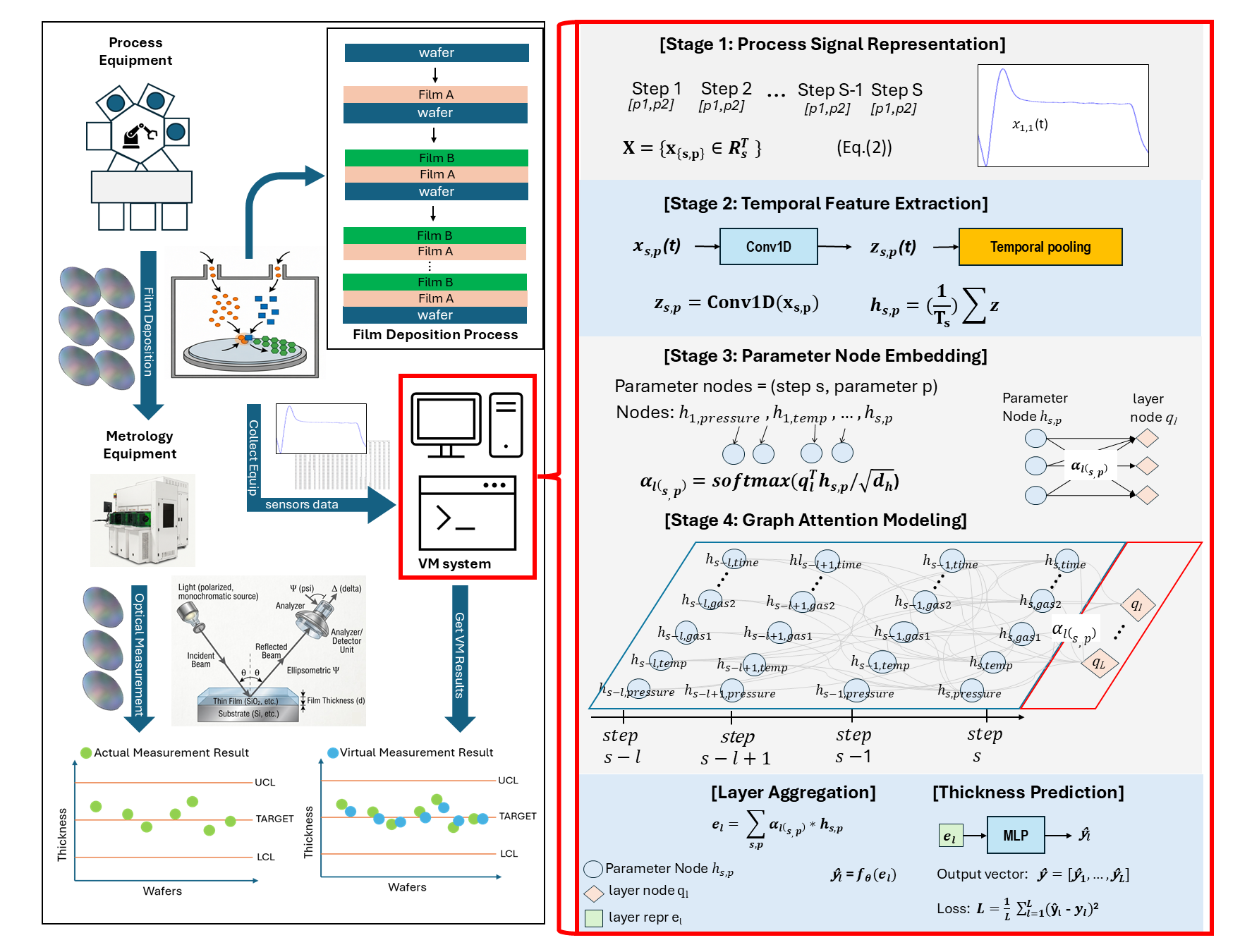}
    \caption{Overall Framework of the Proposed Methodology}
    \label{fig:1 Overall Framework of The Proposed Methodology}
\end{figure*}

To model the relationship between these time-varying process parameters and the resulting film layers, the proposed framework constructs a graph-based representation of the deposition process. 
Each pair of process steps and equipment parameters is represented as a node, and dependencies between parameters and film layers are modeled through a graph attention mechanism.
The overall architecture consists of three components: (1) parameter node embedding, (2) parameter-to-layer dependency modeling via graph attention, and (3) thickness prediction. 
We first formulate the VM prediction problem and then describe the proposed modeling framework.
Figure~\ref{fig:1 Overall Framework of The Proposed Methodology} illustrates the overall framework.

\subsection{Problem Formulation}

Semiconductor film deposition constitutes a highly complex and dynamic process, wherein equipment sensors continuously monitor critical operational parameters—including chamber temperature, pressure, gas flow rates, and radio frequency (RF) power—at high sampling frequencies (e.g., 10 Hz). These real-time temporal variations serve as direct indicators of the internal chamber conditions and the underlying kinetics of film growth. 
Ultimately, the resulting film characteristics, specifically thickness, composition, and uniformity, are governed by the intricate interplay and collective influence of these multifaceted process parameters \cite{semiconprocessfundamentals}.

Consider a film deposition process consisting of $S$ process steps. 
At each step $s$, a set of $P$ equipment parameters is monitored, including variables such as chamber temperature, pressure, and precursor gas flow rates. 
For each parameter $p$ at step $s$, a time-series signal is recorded during wafer processing.
Let $x_{s,p}$ denote the time-series signal corresponding to parameter $p$ measured during process step $s$. 
The signal representation is defined as Eq.~(\ref{eq:signal}):
\begin{equation}
x_{s,p} \in \mathbb{R}^{T_s},
\label{eq:signal}
\end{equation}
where $T_s$ denotes the number of temporal samples in process step $s$.

The complete set of process inputs for a wafer is defined as Eq.~(\ref{eq:input_set}):
\begin{equation}
X = \{ x_{s,p} \in \mathbb{R}^{T_s} \mid s = 1,\dots,S,\; p = 1,\dots,P \},
\label{eq:input_set}
\end{equation}
which represents the collection of multivariate time-series signals observed during the deposition process. 
After wafer processing is completed, metrology tools measure the thickness of deposited film layers. 
Assume that $L$ film layers are deposited during the process. 
The vector of true film thickness measurements for a wafer can be represented as Eq.~(\ref{eq:true_thickness}):
\begin{equation}
y = [y_1, y_2, \ldots, y_L]^\top \in \mathbb{R}^{L},
\label{eq:true_thickness}
\end{equation}
where $y_l$ denotes the measured thickness of the $l$-th film layer.

The goal of VM is to learn a predictive function that maps the process input $X$ to the corresponding film thickness vector. 
This mapping can be expressed as Eq.~(\ref{eq:vm_function}):
\begin{equation}
\hat{y} = f_{\theta}(X),
\label{eq:vm_function}
\end{equation}
where $\hat{y} = [\hat{y}_1, \ldots, \hat{y}_L]^\top \in \mathbb{R}^{L}$ represents the predicted thickness vector and $\theta$ denotes the trainable parameters of the model.
Learning this mapping is challenging for several reasons. 
First, the process input $X$ consists of heterogeneous time-series signals with varying temporal lengths across process steps. 
Second, the influence of process parameters on film thickness may depend on complex interactions among parameters recorded at different stages of the deposition process \cite{HBoffilmdepo}.

To address these challenges, the proposed framework represents each process step–parameter pair $(s,p)$ as a node in a structured representation and extracts temporal features from the corresponding sensor signals. 
These node representations are then used to model relationships between process parameters and deposited film layers for predicting wafer-level thickness measurements.

\subsection{Parameter Node Embedding}

As defined in Eq.~(\ref{eq:signal}), each process parameter is observed as a time-series signal $x_{s,p}$ during process step $s$. 
In semiconductor deposition processes, these equipment traces reflect the temporal evolution of chamber conditions and process states that influence film formation.

Temporal feature extraction from each parameter signal is defined as Eq.~(\ref{eq:conv}):
\begin{equation}
z_{s,p} = \text{Conv1D}(x_{s,p}) \in \mathbb{R}^{T_s \times d_h},
\label{eq:conv}
\end{equation}
where $z_{s,p}$ denotes the temporal feature representation and $d_h$ is the embedding dimension. 
The convolutional filters capture local temporal patterns in equipment traces, enabling the model to encode transient variations in process conditions that influence deposition behavior.

Since the signal length $T_s$ may vary across process steps, temporal aggregation is applied to obtain a fixed-length representation defined as Eq.~(\ref{eq:param_embed}):
\begin{equation}
h_{s,p} =
\frac{1}{T_s}
\sum_{t=1}^{T_s}
z_{s,p}^{(t)}
\in \mathbb{R}^{d_h},
\label{eq:param_embed}
\end{equation}
where $h_{s,p}$ denotes the embedding of the parameter node corresponding to process step $s$ and parameter $p$. 
This representation summarizes the temporal behavior of the corresponding equipment parameter during wafer processing.

The complete set of parameter node embeddings is defined as Eq.~(\ref{eq:param_nodes}):
\begin{equation}
N =
\{ h_{s,p} \mid s = 1,\dots,S,\; p = 1,\dots,P \}
\in \mathbb{R}^{(S \cdot P) \times d_h},
\label{eq:param_nodes}
\end{equation}
where each node in $N$ represents the temporal dynamics of a specific process parameter at a particular process step.

Beyond instantaneous parameter values, thin film deposition is critically governed by their temporal evolution patterns—such as rising, falling, fluctuating, or stabilizing trends—over specific durations. Conv1D are employed to effectively extract these local temporal signatures; for instance, a gradual pressure increase may indicate shifts in deposition kinetics. Given that the duration $T_s$ varies across different process steps, a temporal aggregation (averaging) operation is applied to normalize variable-length time-series features into fixed-length vectors, denoted as $h_{s,p}$, thereby ensuring input consistency for subsequent graph-based models. This vector $h_{s,p}$ encapsulates the comprehensive temporal behavior of a specific parameter within a designated process step, serving as an embedding representation for the corresponding "parameter node" and providing high-level semantic information essential for advanced process modeling.

\subsection{Parameter-to-Layer Dependency Modeling via Graph Attention}

Film thickness in deposition processes is influenced by interactions among multiple equipment parameters across different process stages. 
For example, variations in precursor flow may influence deposition rate, while chamber pressure and temperature affect reaction kinetics and film uniformity. 
Consequently, the relationship between process parameters and film layers can be represented as structured dependencies between equipment conditions and film formation outcomes.

To capture these relationships, the proposed framework constructs a directed bipartite graph consisting of parameter nodes and layer nodes. 
Parameter nodes correspond to the embeddings $N$ defined in Eq.~(\ref{eq:param_nodes}), while each layer node $l$ is represented by a learnable embedding $q_l \in \mathbb{R}^{d_h}$ representing the target film layer.

The attention weight between layer node $l$ and parameter node embedding $h_{s,p}$ is defined as Eq.~(\ref{eq:attn_weight}):
\begin{equation}
\alpha_{l,(s,p)} =
\frac{
\exp\left(q_l^{\top} h_{s,p} / \sqrt{d_h}\right)
}{
\sum_{s'=1}^{S} \sum_{p'=1}^{P}
\exp\left(q_l^{\top} h_{s',p'} / \sqrt{d_h}\right)
},
\label{eq:attn_weight}
\end{equation}
where $\alpha_{l,(s,p)}$ represents the relative importance of parameter $(s,p)$ in determining the thickness of film layer $l$. 
These attention weights allow the model to identify which process parameters contribute most strongly to the formation of each film layer.

The representation of layer node $l$ is computed by aggregating parameter node embeddings according to the attention weights, as defined in Eq.~(\ref{eq:aggregation}):
\begin{equation}
e_l =
\sum_{s=1}^{S} \sum_{p=1}^{P}
\alpha_{l,(s,p)}\, h_{s,p}
\in \mathbb{R}^{d_h},
\label{eq:aggregation}
\end{equation}
where $e_l$ denotes the representation of film layer $l$. 
This aggregation integrates information from multiple equipment parameters and captures the combined influence of chamber conditions on film growth.

From a time-series modeling perspective, the learned attention weights can also be interpreted as directional dependencies between process parameters and film layers. 
Inspired by the concept of Granger causality, these dependencies reflect how historical variations in equipment parameters contribute to film thickness formation during subsequent deposition stages.

The complete set of layer representations is defined as Eq.~(\ref{eq:layer_repr}):
\begin{equation}
E =
\{ e_l \mid l = 1,\dots,L \}
\in \mathbb{R}^{L \times d_h},
\label{eq:layer_repr}
\end{equation}
This stage facilitates the transformation of intricate process data into graph-structured nodes, effectively bridging raw time-series signals and abstract semantic representations. Specifically, each embedding $h_{s,p}$ encapsulates the state of an individual process parameter within a distinct step, while $q_l$ embodies the model's abstract comprehension of the target film layer's desired properties or characteristics. Given the high dimensionality of parameters and the complexity of multi-step deposition sequences, directly modeling the joint influence of all variables presents significant challenges. By decoupling this complexity through the embedding of each "step-parameter" pair as a discrete node, the framework establishes well-defined input units that enable the subsequent graph attention mechanism to efficiently capture interdependencies and refine process modeling.
While film thickness is governed by the synergistic influence of multiple process parameters, their relative significance varies substantially across different contexts. 

The employed attention mechanism autonomously learns and quantifies the contribution of each "step-parameter" pair to the thickness of a specific film layer; for instance, chamber temperature may dominate dielectric layer growth, whereas gas flow rates could be more critical for metal layers. The resulting attention weights, denoted as $\alpha_{l,(s,p)}$, serve as interpretable metrics representing the importance of parameter $(s,p)$ for film layer $l$. This interpretability provides process engineers with actionable insights to identify the critical conditions driving thickness variations, thereby facilitating rigorous root cause analysis and process optimization. Crucially, this capability directly addresses the inherent "black-box" limitation prevalent in existing DL models, which is a primary objective of this study.

\subsection{Film Thickness Prediction}

Given the layer representations $E$, the thickness of each film layer is predicted using a multilayer perceptron (MLP) regression model defined as Eq.~(\ref{eq:pred}):
\begin{equation}
\hat{y}_l = f_\theta(e_l), \quad l = 1,\dots,L,
\label{eq:pred}
\end{equation}
where $f_\theta(\cdot)$ denotes a multilayer perceptron and $\hat{y}_l$ represents the predicted thickness of film layer $l$. 
The predicted thickness vector $\hat{y}$ therefore corresponds to the model output defined in Eq.~(\ref{eq:vm_function}).

The model parameters are optimized by minimizing the mean squared error between predicted and measured thickness values defined as Eq.~(\ref{eq:loss}):
\begin{equation}
\mathcal{L} =
\frac{1}{L}
\sum_{l=1}^{L}
\left( \frac{\hat{y}_l - y_l}{\sigma_l} \right)^2,
\label{eq:loss}
\end{equation}
where $y_l$ denotes the true thickness of the $l$-th film layer defined in Eq.~(\ref{eq:true_thickness}), and $\sigma_l$ is the training-set standard deviation of the $l$-th film used to standardize the targets across different thickness scales. 
All model parameters, including the layer node embeddings $\{q_l\}$ and neural network parameters $\theta$, are optimized jointly through this objective.

\section{Case Study}

This section presents a case study to evaluate the performance and interpretability of the proposed graph attention-based VM framework in a semiconductor deposition process. The objective is to assess the model’s ability to predict film thickness from multivariate equipment sensor data, while examining whether the learned parameter–layer dependencies provide meaningful insight into the underlying process behavior.
The analysis is conducted using industrial deposition data collected from a production environment, where process parameters are recorded across sequential process steps and film thickness measurements are obtained through physical metrology. The study focuses on two aspects: prediction accuracy compared to baseline models, and interpretability of the learned relationships between process parameters and film thickness.

The following subsections describe the dataset, experimental setup, and evaluation results, followed by a discussion of model performance and interpretability.

\subsection{Data Acquisition and Dataset Description}

This study utilizes a dataset consisting of equipment sensor traces collected from a semiconductor film deposition process. The data were obtained from a production environment, where multiple process parameters are recorded throughout wafer processing.
The dataset includes time-series measurements of key process variables, such as chamber temperature, chamber pressure, precursor gas flow rates, and lamp power. These signals are collected at a sampling rate of 10 Hz across multiple process steps, capturing the temporal evolution of process conditions during the deposition sequence. Through a feature selection process guided by domain expertise, 30 critical parameters were identified from an initial pool of over 1,000 equipment variables based on their relevance to film thickness variation.
The dataset consists of approximately 1,000 wafers processed through a multi-layer chemical vapor deposition process, where half of the layers correspond to film type A and the other half to film type B. All wafers are real production wafers obtained from the Intel Advanced Technology Foundry.

Following deposition, film thickness measurements are obtained using spectroscopic ellipsometry at more than 10 spatial locations per wafer to capture intra-wafer non-uniformity. These measurements serve as the ground truth for training and evaluating the VM model.
To prevent data leakage between wafers produced under similar process conditions, the dataset is split at the lot level rather than the individual wafer level. The final dataset contains 804 wafer samples, which are divided into training, validation, and test sets with ratios of 70\%, 15\%, and 15\%, corresponding to 558, 120, and 126 wafers, respectively.

\subsection{Experimental Setup}
\begin{figure}[h]
    \centering
    \includegraphics[width=0.95\columnwidth]{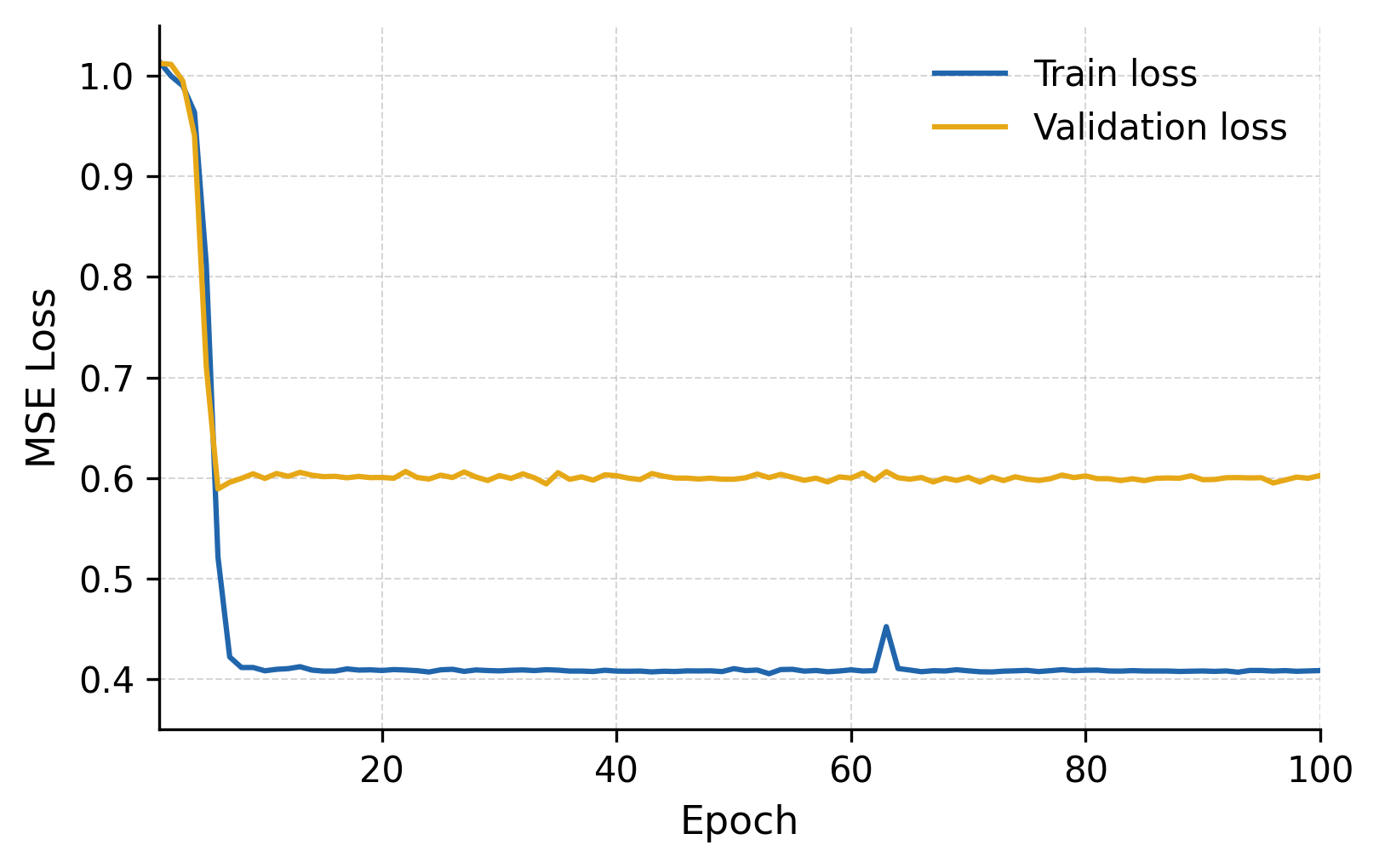}
    \caption{Training and validation loss curves over 100 epochs. Both losses converge rapidly and remain stable thereafter. The gap between training and validation loss reflects the lot-level distribution shift inherent to the lot-based data split strategy, rather than overfitting.}
    \label{fig:loss_curve}
\end{figure}

All experiments were implemented in PyTorch and conducted on a server equipped with an NVIDIA A100 GPU with 80 GB VRAM. The model was trained using the Adam optimizer with a learning rate of $1 \times 10^{-4}$ for 100 epochs and a batch size of 16. The hidden dimension $d_h$ was set to 64, the number of attention heads to 4, and the Conv1D kernel size to 5.
Since the target film thicknesses span different physical scales, each target is standardized independently using the mean and standard deviation computed from the training set, as defined in Eq.~(\ref{eq:loss}). The model checkpoint with the lowest validation loss is selected as the final model.
Figure~\ref{fig:loss_curve} shows the training and validation loss curves over 100 epochs. The loss decreases rapidly during the initial training phase and stabilizes thereafter, indicating effective convergence.
Model performance is evaluated using the coefficient of determination ($R^2$), computed for each film on the original scale as defined in Eq.~(\ref{eq:r2}). 
\begin{equation}
    R^2_l = 1 - \frac{\sum_{i=1}^{N} (y_{i,l} - \hat{y}_{i,l})^2}{\sum_{i=1}^{N} (y_{i,l} - \bar{y}_l)^2}
    \label{eq:r2}
\end{equation}
This metric is scale-independent and allows consistent comparison across films with different thickness ranges.

To evaluate the contribution of the proposed parameter-to-layer attention mechanism, three model configurations are compared. The proposed model uses parameter nodes constructed from all process steps and applies parameter-to-layer attention to model structured dependencies. An ablation variant removes this attention mechanism, eliminating explicit parameter–layer interactions. As a baseline, an MLP is trained on the same input features, consisting of two hidden layers with dimensions 128 and 64 and ReLU activations. The MLP architecture is intentionally kept simple due to the limited size of the training dataset.

\subsection{Results and Discussion}
\begin{figure}[htbp]
\centering
   \includegraphics[width=0.8\columnwidth]{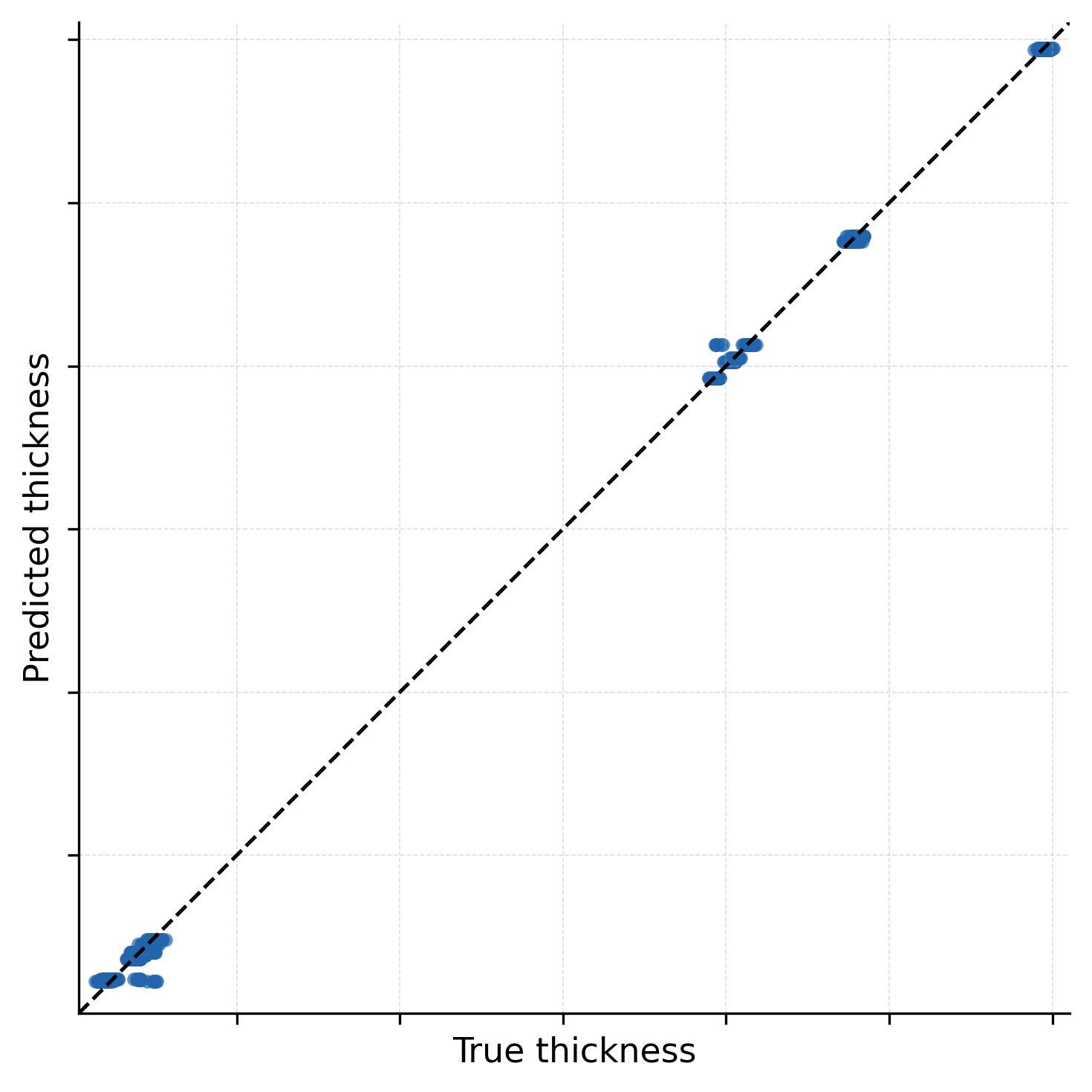}
    \caption{Overall prediction performance on the test set shown as a scatter plot of predicted versus measured thickness values for all target layers. Each point denotes one layer-level sample, and the dashed diagonal line represents ideal prediction. Thickness values are omitted due to confidentiality constraints.}
    \label{fig: Prediction vs True thickness}
\end{figure}

As summarized in Table \ref{tab:3}, the proposed model achieved superior predictive accuracy, attaining an $R^2$ of 0.7227. In comparison, the MLP baseline yielded an $R^2$ of 0.7106, while the ablation variant recorded an $R^2$ of 0.7061. 
The $R^2$ values reported represent the mean across all four Film B targets.
These results demonstrate the enhanced predictive capability of the proposed approach. 
\begin{table}
\caption{Performance Comparison of Model Configurations}\label{tab:3}%
\centering{%
\begin{tabular}{lc} 
\toprule
Configuration & $R^2$ \\
\midrule
\textbf{Proposed model (full)}                          & \textbf{0.7227} \\
Proposed model w/o parameter-to-layer attention & 0.7061 \\
MLP (baseline)                                 & 0.7106 \\
\bottomrule
\end{tabular}
}
\vspace{2pt}
\raggedright
\footnotesize \textit{$^{*}$Mean of the individual $R^2$ scores across the four Film B targets.}
\end{table}
As depicted in Fig.~\ref{fig: Prediction vs True thickness}, the predicted thickness values are closely aligned with the ground-truth metrology measurements across all film types and thickness ranges. The overall concentration of points near the diagonal indicates that the proposed model achieves consistent predictive performance on the test set.


\subsubsection{Attention-Based Interpretability Analysis}

To further investigate the behavior of the proposed model, we analyze the learned attention weights, which quantify the relative importance of process parameters in predicting film thickness for each layer.

\begin{figure}[h]
\centering
   \includegraphics[width=1.0\columnwidth]{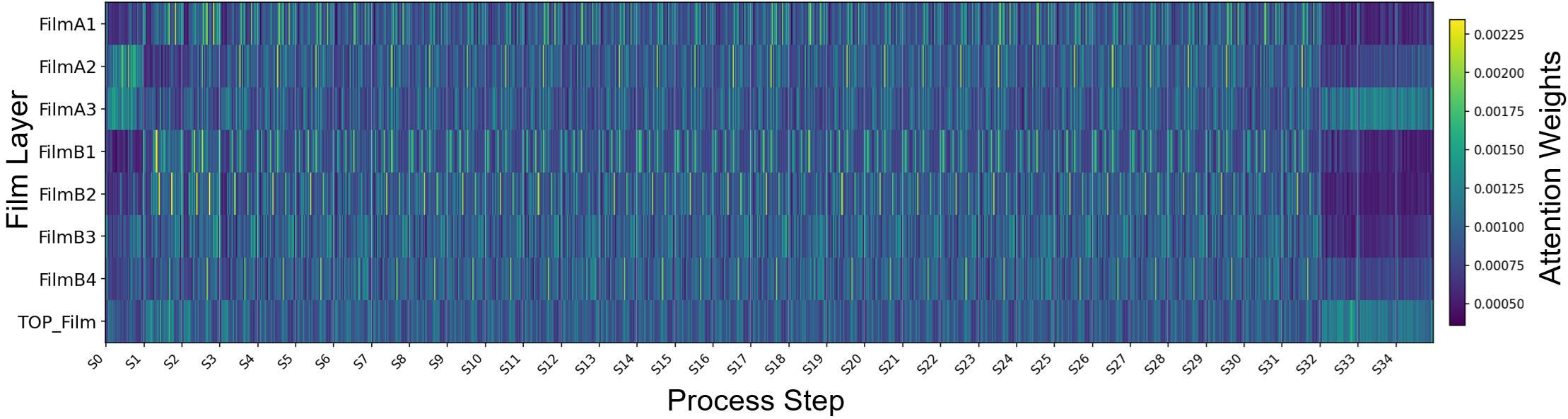}
    \caption{Full heatmap of attention weights averaged over the test set. Each process step contains 30 parameter nodes.}
    \label{fig: Attention Weights Heatmap1}
\end{figure}

Figure~\ref{fig: Attention Weights Heatmap1} presents a heatmap of the attention weights $\alpha_{l,(s,p)}$, where each row corresponds to a film layer and each column represents a process parameter across different process steps. The attention weights are averaged over all samples in the test dataset to obtain a stable and representative parameter–layer dependency structure.
From the full heatmap, several global patterns can be observed. Specific film layers exhibit persistently high attention weights during the initial and final process steps, highlighting their acute sensitivity to boundary conditions. In contrast, all layers display a consistent, repetitive attention pattern throughout the intermediate stages. This uniformity suggests that once chamber temperature and pressure stabilize, their influence on thickness becomes homogeneous across layers, allowing dynamic variables—such as reactive gas flows—to drive variations in attention values. Moreover, these recurring patterns likely reflect the propagation of early-stage parameter effects into later layers. This behavior reflects the cumulative and history-dependent nature of deposition processes, where prior process conditions influence subsequent film growth through mechanisms such as surface conditioning, residual species, and thermal accumulation effects.

\begin{figure}[t]
\centering
   \includegraphics[width=0.99\columnwidth]{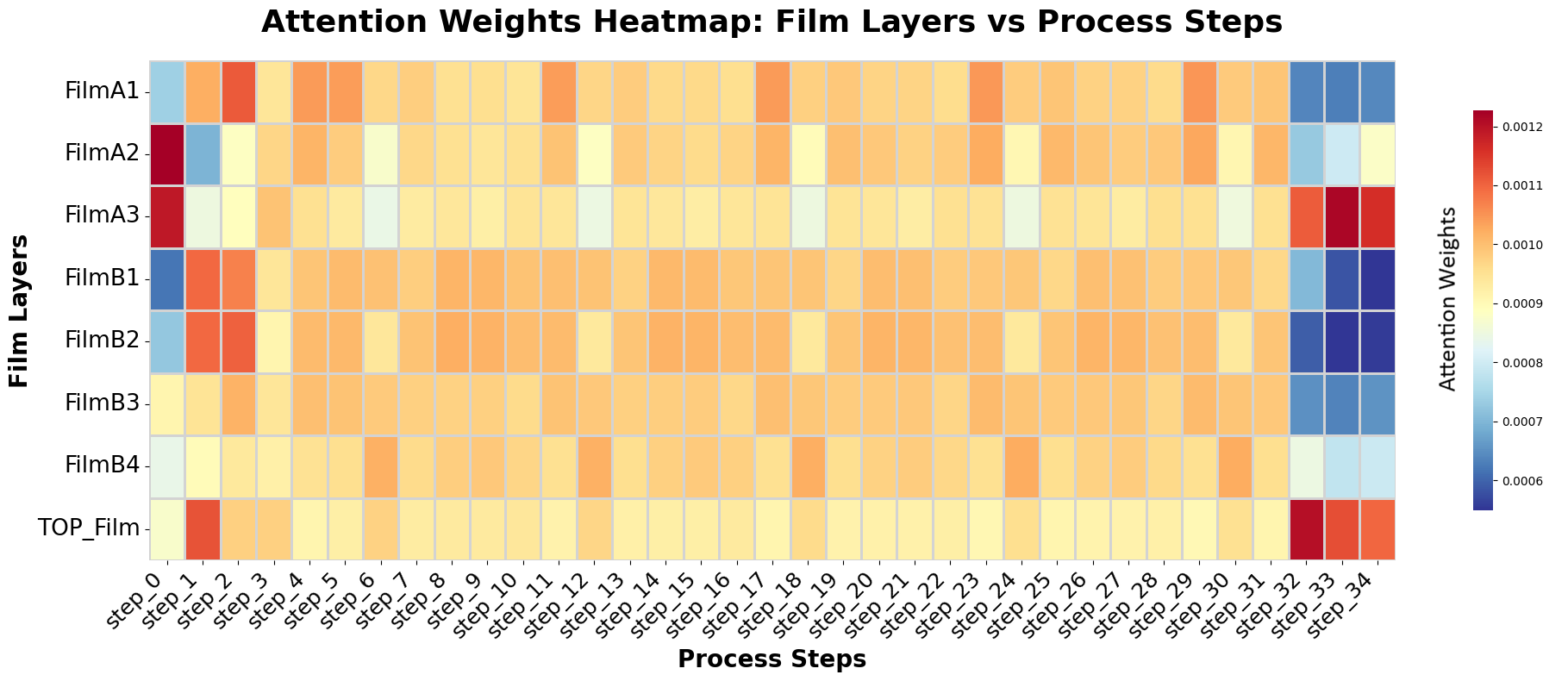}
    \caption{Heatmap of The Top-k Attention Weights}
    \label{fig: Attention Weights Heatmap2}
\end{figure}

Due to the high dimensionality of the process space, where approximately 30 process parameters are recorded across 35 sub-steps, resulting in over 1,000 effective input variables, the full attention map contains complex patterns that are difficult to interpret directly. 
Therefore, we further refine the visualization by focusing on the most influential parameters.
Figure~\ref{fig: Attention Weights Heatmap2} shows a filtered heatmap in which only the top-$k$ attention weights are retained for each layer, while the remaining values are masked. The filtered heatmap reveals that each layer is primarily influenced by a small subset of process parameters, rather than the entire parameter space. This sparsity indicates that film thickness formation is governed by a limited number of critical process factors at each stage.
The heatmap reveals that the model effectively identifies critical process dynamics, assigning peak attention weights to the first and last three steps associated with temperature and pressure ramping. This confirms the model's sensitivity to abrupt process transitions. Distinct layer-specific behaviors are evident: Film B layers correlate strongly only with the initial ramping phase, showing negligible response to the final steps. In contrast, while most Film A layers respond to initial conditions, Film A3 and the Top Film exhibit significant sensitivity to the terminal ramping phase. Notably, periodic patterns observed across all layers during intermediate steps, which aligned with the findings in full heatmap.
Additionally, the variation in dominant parameters across layers highlights layer-specific process sensitivity, which is consistent with the fact that different layers may involve distinct deposition regimes or material interactions.


\begin{figure*}[h]
\centering
   \includegraphics[width=0.98\textwidth]{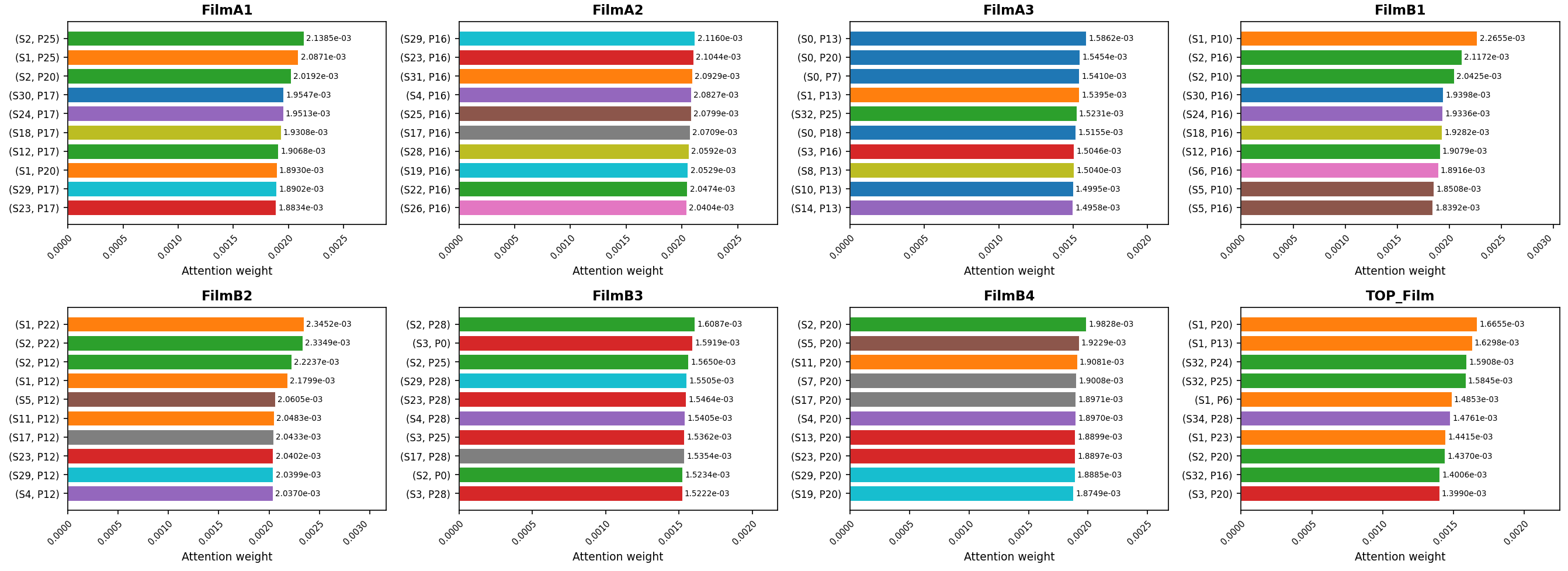}
    \caption{Bar Charts Of The Top-$k$ ($k$=10) Parameters For Selected Layers}
    \label{fig: Attention Weights Heatmap3}
\end{figure*}

To further examine these relationships, Figure~\ref{fig: Attention Weights Heatmap3} presents bar charts of the top-$k$ parameters for selected layers.
As illustrated in the bar chart, all film layers exhibit strong causality with temperature-related parameters. Moreover, the initial processing steps exert the most significant influence across all film layers except film A2. The variations among the ten listed parameters are negligible. Consistent with the top-k attention weight heatmap results and theoretical expectations derived from the film deposition process, the top film layer additionally demonstrates pronounced causality with the terminal processing steps.
These results demonstrate that different layers are governed by distinct physical mechanisms, and the model successfully captures these variations through the learned attention weights.
From a time-series modeling perspective, the learned attention weights can be interpreted as directional dependencies between process parameters and film layers. Inspired by the concept of Granger causality, these dependencies capture how historical variations in equipment parameters contribute to film thickness formation in subsequent deposition stages.

This analysis bridges the gap between data-driven modeling and process understanding, providing actionable insights for process monitoring and optimization in semiconductor manufacturing.
Overall, this analysis demonstrates that the proposed graph attention-based VM framework not only achieves accurate predictions but also provides interpretable insights into the underlying process dynamics.

\section{Concluding Remarks and Future Work}

In this study, a graph attention-based VM framework was developed for film thickness prediction in semiconductor deposition processes. 
The proposed model captures structured dependencies between process parameters and film layers, enabling improved predictive performance compared to baseline approaches.
A key contribution of this work lies in the interpretability of the parameter-to-layer attention mechanism. The learned attention weights provide a quantitative measure of the relative importance of process parameters for each film layer, offering insight into the underlying process dynamics. 
The observed parameter–layer relationships are consistent with physical intuition, reflecting the roles of temperature, pressure, and gas flow in governing deposition kinetics and film growth behavior.
From a time-series perspective, the attention mechanism captures directional dependencies between process parameters and film layers. 
The interpretability of the model enables the identification and ranking of critical parameters and process steps that influence film thickness.
This facilitates more informed root cause analysis and supports targeted process optimization.

Several directions remain for future work. First, the validation of learned parameter–layer relationships against physical and chemical principles should be pursued through close collaboration with domain experts. 
Second, extending the framework to capture within-wafer thickness variation and spatial non-uniformity would improve its applicability in real manufacturing environments. 
Finally, incorporating additional process knowledge or physics-informed constraints may further enhance model robustness and generalization under process drift and variability.


\section*{Acknowledgments}
The authors acknowledge the Intel Foundry OTF Campus for providing access to advanced semiconductor manufacturing production data used in this study.



\bibliographystyle{asmeconf}  

\begin{thebibliography}{10}
\newcommand{\enquote}[1]{``#1''}
\providecommand{\url}[1]{\texttt{#1}}
\providecommand{\urlprefix}{URL }
\expandafter\ifx\csname urlstyle\endcsname\relax
  \providecommand{\doi}[1]{DOI \discretionary{}{}{}#1}\else
  \providecommand{\doi}{DOI \discretionary{}{}{}\begingroup \urlstyle{rm}\Url}\fi
\providecommand{\eprint}[2][]{\urlprefix\url{#1#2}}

\bibitem{siliconminds}
Talati, Dhruvitkumar.
\newblock \enquote{Silicon minds: The rise of AI-powered chips.}  (2021).

\bibitem{developmenttrendsofsemicon}
Qiu, Zekun, Shen, Xianao and Zhao, Zirui.
\newblock \enquote{Development trends and prospects of semiconductor devices and technology.}
\newblock \textit{Development} Vol.~81 (2024).

\bibitem{nanoscalechallenge}
Srivastava, Sumit, Jaiswal, Abhinav and Khan, Arman.
\newblock \enquote{Nanoscale technologies: design challenges and advancements.}
\newblock \textit{Semiconductor Nanoscale Devices: Materials and Design Challenges}.
\newblock Bentham Science Publishers (2025): pp. 1--26.

\bibitem{lee2026generative}
Lee, Suk~Ki, Stone, Ronnie~FP, Gao, Max, Zhang, Wenlong, Sha, Zhenghui and Ko, Hyunwoong.
\newblock \enquote{Generative Model Predictive Control in Manufacturing Processes: A Review.}
\newblock \textit{arXiv preprint arXiv:2511.17865}  (2025).

\bibitem{ThinFilmreview}
Sakthinathan, Subramanian, Meenakshi, Ganesh~Abinaya, Vinothini, Sivaramakrishnan, Yu, Chung-Lun, Chen, Ching-Lung, Chiu, Te-Wei and Vittayakorn, Naratip.
\newblock \enquote{A review of thin-film growth, properties, applications, and future prospects.}
\newblock \textit{Processes} Vol.~13 No.~2 (2025): p. 587.

\bibitem{integratedcircuitfabrication}
Plummer, James~D and Griffin, Peter~B.
\newblock \textit{Integrated Circuit Fabrication: Science and Technology}.
\newblock Cambridge University Press (2023).

\bibitem{metrologyfornextgensemicon}
Orji, Ndubuisi~G, Badaroglu, Mustafa, Barnes, Bryan~M, Beitia, Carlos, Bunday, Benjamin~D, Celano, Umberto, Kline, Regis~J, Neisser, Mark, Obeng, Yaw and Vladar, AE.
\newblock \enquote{Metrology for the next generation of semiconductor devices.}
\newblock \textit{Nature electronics} Vol.~1 No.~10 (2018): pp. 532--547.

\bibitem{metrorequirement}
Bunday, Benjamin, Bello, AF, Solecky, Eric and Vaid, Alok.
\newblock \enquote{7/5nm logic manufacturing capabilities and requirements of metrology.}
\newblock \textit{Metrology, Inspection, and Process Control for Microlithography XXXII}, Vol. 10585: pp. 81--124. 2018. SPIE.

\bibitem{exploringMLpredict}
Chen, Ying-Lin, Sacchi, Sara, Dey, Bappaditya, Blanco, Victor, Halder, Sandip, Leray, Philippe and De~Gendt, Stefan.
\newblock \enquote{Exploring machine learning for semiconductor process optimization: A systematic review.}
\newblock \textit{IEEE Transactions on Artificial Intelligence} Vol.~5 No.~12 (2024): pp. 5969--5989.

\bibitem{approachVM}
Khan, Aftab~A, Moyne, James~R and Tilbury, Dawn~M.
\newblock \enquote{An approach for factory-wide control utilizing virtual metrology.}
\newblock \textit{IEEE Transactions on semiconductor Manufacturing} Vol.~20 No.~4 (2007): pp. 364--375.

\bibitem{VMdevelopmentforSIN}
Roh, Hyun-Joon, Ryu, Sangwon, Jang, Yunchang, Kim, Nam-Kyun, Jin, Younggil, Park, Seolhye and Kim, Gon-Ho.
\newblock \enquote{Development of the virtual metrology for the nitride thickness in multi-layer plasma-enhanced chemical vapor deposition using plasma-information variables.}
\newblock \textit{IEEE Transactions on Semiconductor Manufacturing} Vol.~31 No.~2 (2018): pp. 232--241.

\bibitem{Review_AI_for_OP_Met}
Xu, Weiwang, Zhang, Houdao, Ji, Lingjing and Li, Zhongyu.
\newblock \enquote{AI-Powered Next-Generation Technology for Semiconductor Optical Metrology: A Review.}
\newblock \textit{Micromachines} Vol.~16 No.~8 (2025).
\newblock \doi{10.3390/mi16080838}.

\bibitem{semiconprocessfundamentals}
El-Kareh, Badih and Hutter, Lou~N.
\newblock \textit{Fundamentals of semiconductor processing technology}.
\newblock Springer Science \& Business Media (2012).

\bibitem{decisionVM}
Chien, Chen-Fu, Hung, Wei-Tse, Pan, Chin-Wei and Van~Nguyen, Tran~Hong.
\newblock \enquote{Decision-based virtual metrology for advanced process control to empower smart production and an empirical study for semiconductor manufacturing.}
\newblock \textit{Computers \& Industrial Engineering} Vol. 169 (2022): p. 108245.

\bibitem{CDSEM}
Postek, Michael~T and Vlad{\'a}r, Andr{\'a}s~E.
\newblock \enquote{Critical-dimension metrology and the scanning electron microscope.}
\newblock \textit{Handbook of Silicon Semiconductor Metrology}.
\newblock CRC Press (2001): pp. 244--275.

\bibitem{SE}
Politano, Grazia~Giuseppina and Versace, Carlo.
\newblock \enquote{Spectroscopic ellipsometry: advancements, applications and future prospects in optical characterization.}
\newblock \textit{Spectroscopy Journal} Vol.~1 No.~3 (2023): pp. 163--181.

\bibitem{metrology}
Bunday, Benjamin and Orji, George.
\newblock \enquote{Metrology.}
\newblock \textit{2021 IEEE International Roadmap for Devices and Systems Outbriefs}: pp. 01--68. 2021. IEEE.

\bibitem{datadrivenmetro}
Schneider, Linda-Sophie, Krauss, Patrick, Schiering, Nadine, Syben, Christopher, Schielein, Richard and Maier, Andreas.
\newblock \enquote{Data-driven modeling in metrology--A short introduction, current developments and future perspectives.}
\newblock \textit{tm-Technisches Messen} Vol.~91 No.~9 (2024): pp. 480--503.

\bibitem{VMforsemiconmfg}
Bertorelle, Nicola.
\newblock \enquote{Virtual metrology for semiconductor manufacturing applications.}  .

\bibitem{reviewregandpred}
Chen, Hsuan-Yu and Chen, Chiachung.
\newblock \enquote{Review of Applications of Regression and Predictive Modeling in Wafer Manufacturing.}
\newblock \textit{Electronics} Vol.~14 No.~20 (2025): p. 4083.

\bibitem{cnnattentionforpred}
Zhang, Pengju, Pan, Hao, Chen, Chen, Jing, Yiming and Liu, Ding.
\newblock \enquote{CNN--BiLSTM--Attention-Based Hybrid-Driven Modeling for Diameter Prediction of Czochralski Silicon Single Crystals.}
\newblock \textit{Crystals (2073-4352)} Vol.~16 No.~1 (2026).

\bibitem{lee2024amtransformer}
Lee, Suk~Ki and Ko, Hyunwoong.
\newblock \enquote{AMTransformer: A Koopman theory-based transformer for learning additive manufacturing dynamics in laser processes.}
\newblock \textit{International Journal of AI for Materials and Design} Vol.~1 No.~2 (2024): pp. 76--91.

\bibitem{GNNVMPVD}
Zhou, Longfei, Jin, Dong, Chen, Shuangwu, Yang, Jian and Xie, Jian.
\newblock \enquote{Virtual Metrology Based on Graph Convolutional Neural Network for Semiconductor PVD Process.}
\newblock \textit{2024 5th International Conference on Artificial Intelligence and Electromechanical Automation (AIEA)}: pp. 1053--1058. 2024. IEEE.

\bibitem{SukkiLee31122026}
Lee, Suk~Ki, Kim, Wonah, Lee, Sungbeom, Park, Jeonghyeon, Chun, Sejin, Yeung, Ho and Ko, Hyunwoong.
\newblock \enquote{Graph attention-based dynamical and causal spatiotemporal learning for anomaly detection in additive manufacturing.}
\newblock \textit{Virtual and Physical Prototyping} Vol.~21 No.~1 (2026): p. e2611194.
\newblock \doi{10.1080/17452759.2025.2611194}.

\bibitem{multistageprocessdiag}
Choi, Jongwon and Kim, Seoung~Bum.
\newblock \enquote{Multi-stage process diagnosis networks in semiconductor manufacturing.}
\newblock \textit{IEEE Access} Vol.~12 (2024): pp. 39495--39504.

\bibitem{lehnert2024xplainable}
Lehnert, Alexander, Gawantka, Falko, During, Jonas, Just, Franz and Reichenbach, Marc.
\newblock \enquote{XplAInable: Explainable AI Smoke Detection at the Edge.}
\newblock \textit{Big Data and Cognitive Computing} Vol.~8 No.~5 (2024): p.~50.

\bibitem{HBoffilmdepo}
Seshan, Krishna.
\newblock \textit{Handbook of thin film deposition}.
\newblock William Andrew (2012).

\end{thebibliography}

\end{document}